# convoHER2: A Deep Neural Network for Multi-Stage Classification of HER2 Breast Cancer

M. F. Mridha, *Fellow,* Md. Kishor Morol, Md. Asraf Ali, and Md Sakib Hossain Shovon

*Abstract*— Generally, human epidermal growth factor 2 (HER2) breast cancer is more aggressive than other kinds of breast cancer. Currently, HER2 breast cancer is detected using expensive medical tests are most expensive. Therefore, the aim of this study was to develop a computational model named convoHER2 for detecting HER2 breast cancer with image data using convolution neural network (CNN). Hematoxylin and eosin (H&E) and immunohistochemical (IHC) stained images has been used as raw data from the Bayesian information criterion (BIC) benchmark dataset. This dataset consists of 4873 images of H&E and IHC. Among all images of the dataset, 3896 and 977 images are applied to train and test the convoHER2 model, respectively. As all the images are in high resolution, we resize them so that we can feed them in our convoHER2 model. The cancerous samples images are classified into four classes based on the stage of the cancer (0+, 1+, 2+, 3+). The convoHER2 model is able to detect HER2 cancer and its grade with accuracy 85% and 88% using H&E images and IHC images, respectively. The outcomes of this study determined that the HER2 cancer detecting rates of the convoHER2 model are much enough to provide better diagnosis to the patient for recovering their HER2 breast cancer in future.

## I. INTRODUCTION

CANCER is a fatal disease in which cells of the body grow abnormally and spread uncontrollably to the other parts of the body. Among several types of cancer, breast cancer is the most significant global health agitation [1]. The term breast cancer refers to a malignant tumor of the breast cells. It is the most repeatedly diagnosed disease in women worldwide. In the United States, breast cancer is responsible for 30% of female cancer have been estimated in 2020 [2]. Each individual cancer has distinct characteristics. Based on characteristics, some significant subtypes of breast cancers are luminal A, luminal B, human epidermal growth factor receptor 2 (HER2), basal, and normal types [3]. HER2 is the most lethal subtype among them and the most frequently diagnosed cancer in women.

Appropriate diagnosis in the early stage and treatment can remarkably reduce the mortality rate from breast cancer (BC) [4]. According to the regulation of the college of American Pathologists in breast cancer, HER2 quantification must be tested routinely [5]. In addition, for diagnosed breast cancer, it is necessary to determine its status and to make decisions precisely by analyzing the expression of specific proteins, such as HER2 [6]. Usually, pathologists use the hematoxylin & eosin (H&E/HE) method to identify various morphological information such as the shape, pattern and structure of the cells and tissues that help to diagnose cancer. Besides, the immunohistochemical (IHC) staining process is used to determine whether or not the cancer cells have receptors. It characterizes different tumor subtypes and provides information about tumors post-treatment [7]. So that patients can be able to make decisions precisely at the right time and reduce the possibility of the hazardous effect of breast cancer by taking treatment properly. The routine evaluation of HER2 status is driven by IHC artifices [8]. Pathologists examine IHC-stained slides to finalize HER2 status [5]. The present directions revised by the College of American Pathologists/American Society of Clinical Oncology (CAP/ASCO) define a positivity score of HER2 between 0 - 3+, derived empirically through optical perception among IHC-stain. HER2-negative (HER2-) are classified as scoring 0 or 1+. HER2-positive (HER2+) is defined with a score of 3+. In addition, a score of 2+ refers to the ambiguous expression of HER2 [5]. HER2+ tumors tend to expand rapidly compared to HER2- tumors, though HER2+ cancers benefit more from targeted therapy with HER2-targeting medications [9-10]. Trastuzumab, lapatinib, etc. medicines are available to reduce the mortality rate, though those drugs are too expensive; in some cases, not only useless but also harmful [11].

Moreover, through the situ hybridization (FISH) test, HER2 overexpression can be placed. However, it will be accurate but expensive. On the other hand, HER2 neu immunostaining can be tested, which is inexpensive but less accurate [3]. Hence, we need to find an approach to mitigate all these issues concerning cost, time, easiness, and precise decision making. Here, the term comes 'Deep Learning.'

Deep learning is becoming robust and helpful day by day in this era. In digital pathology, the methodology was successfully employed to analyze lymph nodes. During the CAMLYON16 challenge, the deep learning model performed surprisingly very similarly compared with the performance of a human pathologist [13]. The medical imaging community's interest in using these methods to increase cancer screening accuracy is driven by the fast growth of machine learning, particularly deep learning. The deep-learning algorithm can be applied to support the decision-making process in cancer research. Convolution neural network (CNN) is used to categorize breast cancer from the image dataset. The images are categorized using a convolutional neural network. It uses the breast cancer dataset's images as its input. The images are sent to CNN along with the corresponding weights as input. To minimize the losses and to improve performance, the consequences are adjusted Convolution, pooling, ReLU, and fully linked that make up CNN. A feature map is used in the convolution layer to extract the provided image's features and compress the original image. The image's dimensions are decreased by using a pooling layer. When utilized as an activation function, the ReLU layer determines whether or not the value of the activation function falls within a specific range. The last layer of the model is fully connected. The probability of each output class is calculated by combining the results from all layers and using the Softmax function. In breast cancer research, deep learning has significantly contributed over the last decade to the image domain [13-15]. That is why we decided to use deep learning approach to detect HER2 status from IHC & HE medical test slides.

As we mentioned earlier about the instances scoring 0+ or 1+ are categorized as HER2-negative, while cases scoring 3+ are categorized as HER2-positive. Cases with a score of 2 or above are considered borderline. Currently, there are no morphological markers on H & E slides that reliably predict the HER2 status, with the exception of very well-differentiated tumors with low nuclear/cytoplasm area ratios, which are frequently hormonally driven and consequently typically negative for HER2. In the case that findings are ambiguous, it is usual practice to do a further immunohistochemical and molecular examination. Despite the efficacy of immunohistochemistry (IHC), the additional expense and time spent on these tests may be eliminated if all the information required to infer the HER2 status could be derived from H&E whole slide images (WSI) as a preliminary indication of the IHC result. In terms of supervised algorithms in particular, datasets are a crucial part of visual translation. But in fact, very limited datasets are available for H&E and IHC HER2+ identification. Again, such datasets are pretty hard to get and demand large spaces for preprocessing operations to prepare for our study work. The BCI-dataset [16] comprises 9740 pathological sliced frames of 1024*1024 pixels. The IHC and HE findings of various patients contained four categories: 0, 1+, 2+, and 3+.

So far, deep learning is being utilized in different fields for image recognition. In addition, the convolutional neural network (CNN) has achieved state-of-the-art image classification performance compared to other deep learning architectures. The network can process the original picture without laborious image preparation. This paper presents a modified deep learning architecture based on ImageNet weight with inception-V3 [17] named convoHER2. We have used the BCI dataset [16] which contains 3896 training and 977 test H&E and IHC images. We trained our model on H&E and IHC images. There has not been significant research applying deep learning techniques on H&E and IHC images to predict HER2 status and detect HER2 multiple-stage score classification of breast cancer. Eventually, our research direction motivates us to sight this area of research. However, our models outperformed effective results in terms of accuracy.

The paper is organized as follows. In section II, we discuss the background of our work and the works of others in the same field that inspired and helped us in some way. In section III, we discuss the proposed method for this work. Section IV discusses the implementation of the methodology and discusses the experimental results. And finally, Section V has all the conclusion and possible director for future improvements.

## II. RELATED WORKS

Day by day research has been taking place in the field of digital pathology to automate the whole process. Computer-assisted image analysis systems have been developed to assist human pathologists to achieve accurate results. New improving deep learning techniques are being developed to detect breast cancerous (HER2) cells in the early stages, to minimize the cost and reduce the rate of mortality.

Manual method to detect HER2 and its status from HE-stain and IHC-stain are done by expert pathologists with the help of expensive microscope [5, 7]. However, such methods to detect HER2 status are subject to error-prone, as it requires human interpretation [18]. Hence, researchers around the world have developed a variety of automated methods to classify HER2 status from IHC-stain and as well as HE-stain. Moreover, HER2 status was classified from MRI and Ultrasound images [19, 20]. The method in [19] employed Support Vector Machine (SVM) to detect HER2 status from MRIs. The dataset contained an IHC-determined HER2 score of 0-1+ or 3+. A SVM module produced radionics signature from the input MRI with annotated tumour region that accurately predicts HER2 status (accuracy of 79.5% and 78.3% while train and test). However, generalization of MRI data and multicenter validation studies were absent in this study. Moreover, this method failed to surpass pathologists accuracy. DenseNet backed deep learning framework was presented in [20]. The framework utilized 144 cross-section ultrasound images for train and test. The architecture yielded an AUC of 0.87 (accuracy=85.19%, sensitivity=75.53%) during training and 0.84 (accuracy=80.56%, sensitivity=72.73%) on the test set, respectively. However, no benchmark dataset contains ultrasound pictures of the breast tumour location. Hence, further study was focused on detecting HER2 from tissue slices to mimic the typical HER2 status detection method better in [21-27].

Images of HE-stain were utilized for HER2 status detection in [21 - 23]. The framework in [21] contains UNet to detect the positions of nuclei in WSI patches of HE-stain. In addition, it employed a cascade of CNN architecture to classify HER2- or HER2+. The proposed framework achieved AUC value of 0.82 in the validation set of Warwick dataset [28]. Also, 0.76 AUC in an independent dataset named TCGA-BRCA.

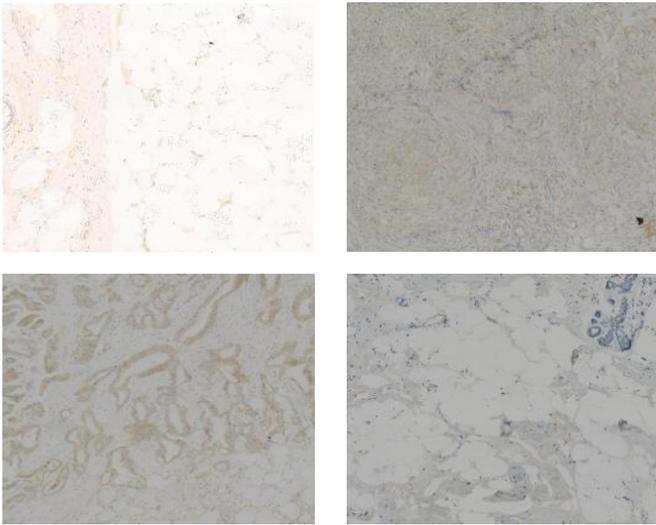

Fig. 1. Sample Images of IHC. Top left is HER2(0), Top right is HER2(1+), Bottom left is HER2(2+) and Bottom right is HER2 (3+).

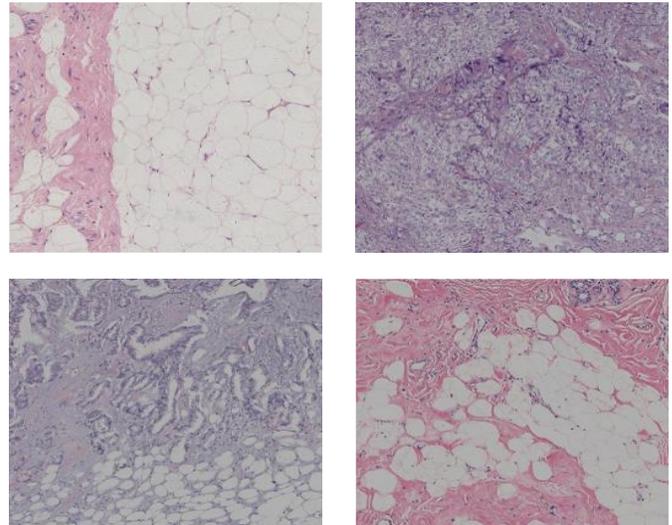

Fig. 2 Sample Images of H&E. Top left is HER2(0), Top right is HER2(1+), Bottom left is HER2(2+) and Bottom right is HER2 (3+).

However, the proposed method does not report patch-level and slide-level AUC separately. In addition, the system has trouble finding HER2+ cells with a score of 2+ (0.73 AUC). In [22], the authors suggested the Inception v3 model with a fine-tuned last layer to classify HER2 status from annotated HE-stained WSIs. The classifier accurately distinguishes HER2 scores and detects trastuzumab reaction in BC treatment. The methodology while training, obtained 0.88 AUC for tile-level and 0.88 AUC and 0.90 AUC at slide-level for classifying HER2-/+ class. The model was validated on the TCGA-BRCA cohort, which obtained AUCs were 0.81 and 0.65 at slide levels and tile levels, respectively. However, the validation of the model was done by exploiting the test dataset, which yielded better AUC results. W. Lu in [23] proposed GNN based framework to predict DAB density from HE stained WSIs and HER2 score from generated DAB density. The mentioned architecture obtains 0.75 AUC in the TCGA-BRCA test set and 0.78 and 0.80 AUCs in the HER2C and Nott-HER2 datasets. However, HER2+, score 2+, was avoided while testing the model.

Apart from detecting HER2 status from HE-stain, it was more accurately obtained by utilizing IHC-stain images [24-27]. Machine learning-based approaches were proposed in [24, 25] to detect HER2 status from IHC-stain. In [24], the method utilized colour features to classify HER2 status scores from labelled IHC images of the Warwick dataset. The method includes patch-level classification and whole slide image (WSI) classification of HER2 status. The proposed method achieved 90.20% and 94.12% classification accuracy in patch-level with MLP model along with HSV generated colour feature and WSI-level with SVM utilizing HSV feature. Although other types of machine learning models, namely KNN and Decision Tree, were evaluated, their performance is not worth mentioning. The logistic regression-based model in [25] achieved HER2 status score classification accuracy of 93%. However, this reported accuracy in [24, 25] was only obtained from the training phase.

Since IHC-stained WSIs for different HER2 scores have unique features in terms of colour and structural pattern of the cell membrane and nucleus [26]. Hence, the CNN-based framework can better classify IHC-stained images based on HER2 score. CNN models are mainly designed to learn features from images to be employed for various image classification problems [29].

The authors in [26-27] solved the classification problem of HER2 status from IHC-stained images using CNN-based deep learning architecture.

The architecture in [26] incorporated convolutional, deconvolution and TLSTM modules. The architecture was trained on exploited IHC images of the Warwick dataset. The framework achieved an accuracy of 98.33% on the held-out dataset. Hence, CNN based framework obtained significantly better accuracy than machine learning-based methods. However, the method scores HER2 status based on segmented cell membrane and nucleus, which is computationally complex and inefficient. Again, the proposed method only considers patch-level accuracy and lacks WSI-level accuracy. Both the patch and WSI level accuracy were reported in [27]. The framework was VGG19 based, that only considers three classes, which are positive (score 3+), equivocal (scoring 2+), and negative (score 0/1+). However, the space complexity of VGG19 architecture is severe. Also, the authors suggested labelling each patch with their respected HER2 score instead of the WSI level score to improve accuracy.

For the purpose of producing an IHC label, Oliveira et al. [30] introduced multiple instance learning (MIL) methodology. The methodology incorporated CNN namely HASHI algorithm. The CNN block processed HE-stain image patches independently. HER2SC slides was fed to train the architecture and tested on subsets of HER2SC and TCGA-TCIA-BRCA (BRCA) slides. Classification accuracy of 83.3% and 53.8% was reported in the HER2SC validation data and the BRCA validation data.

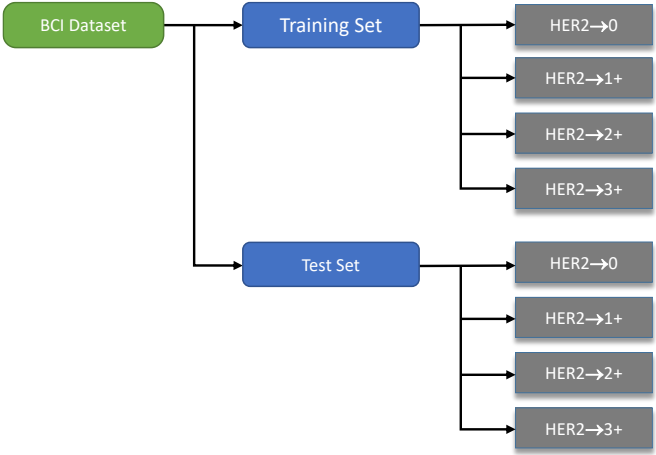

Fig. 3. Distribution of the dataset

Liu et al. in [16] proposed a sophisticated dataset called BCI, containing 4870 image patches of HE-stain. Also, it contains 4870 image patches of IHC-stain of the same HE-stain with an accurate label of HER2 status score in each image. The HE-stain and IHC-stain images of the BCI dataset are structurally aligned. The dataset is distributed with 240 patches of IHC score 0, 1153 patches of IHC score 1+, 2142 patches of IHC score 2+, and 1335 patches of IHC score 3+. The dataset was used to generate IHC-stain images from He-stain images in [17]. In that case, a generative model called pyramid pix2pix was proposed. The results obtained were 21.160 for Peak Signal to Noise Ratio (PSNR) and 0.477 for Structural Similarity (SSIM) matric in the BCI dataset and PSNR of 12.91 and SSIM of 0.278 in the LLVIP dataset, which was the current best compared to other image generative frameworks namely, pix2pix, pix2pixHD, and cycleGAN. Based on the pathologist's evaluation, the generated images achieved an average accuracy of only 38.75%. The results indicate that such a generative method is not suitable yet to predict HER2 status, so we need to focus on the classification problem of IHC stain images.

Hence, we propose to utilize transfer-learning methodology in end-to-end deep learning architectures on a

TABLE I
MODEL SUMMARY OF COVOHER2

| Layer(type) | Output Shape | Param # |
|---|---|---|
| inception_v3 (Functional) | (None, 2048) | 21802784 |
| flatten (Flatten) | (None, 2048) | 0 |
| batch_normalization_94 (BatchNormalization) | (None, 2048) | 8192 |
| dense (Dense) | (None, 2048) | 4196352 |
| batch_normalization_95 (BatchNormalization) | (None, 2048) | 8192 |
| dense_1 (Dense) | (None, 1536) | 3147264 |
| batch_normalization_96 (BatchNormalization) | (None, 1536) | 6144 |
| dense_2 (Dense) | (None, 1536) | 2360832 |
| batch_normalization_97 (BatchNormalization) | (None, 1536) | 6144 |
| dense_3 (Dense) | (None, 4) | 6148 |

Total params: 31,542,052
Trainable params: 9,724,932
Non-trainable params: 21,817,120

well-prepared dataset like BCI and establish robustness to detect HER2 score from both HE-stain and IHC-stained images.

.

### III. METHODOLOGY

#### A. Dataset Description

The breast cancer immunohistochemical (BCI) benchmark dataset was applied in this research. This benchmark aims to directly synthesize immunohistochemical techniques (IHC) data with HE-stain. The dataset includes 9740 images that represent a range of HER2 scores. The resolution of the images was 1024 × 1024 which was collected from 51 patients. The images are categorized into four stages of HER2 cancer.

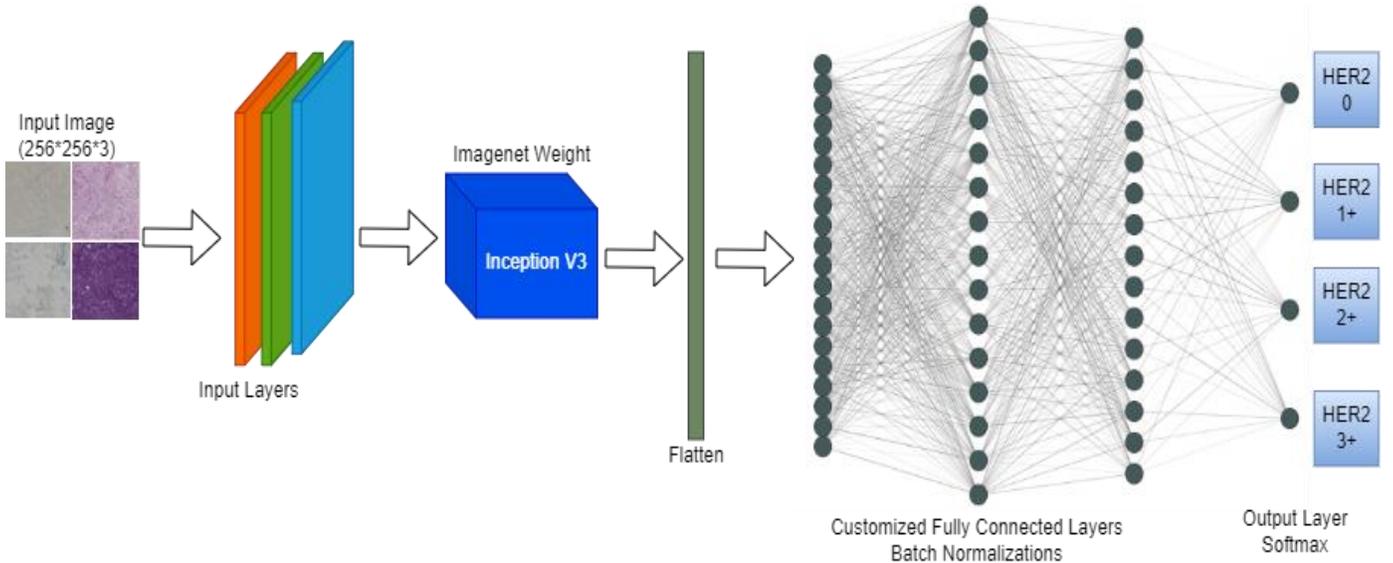

Fig. 4. Proposed architecture of convoHER2

The H&E and IHC datasets were chosen due to their availability and significance. There are a few other processes to determine if breast cancer has developed or not. But IHC and H&E are the most practical and precise ways to determine the exact stage (0, 1, 2, or 3). But the connection between IHC and H&E is what led to the inclusion of two medical test image datasets in this study. On the one hand, it is highly challenging to determine the stage of breast cancer using B&E test images, and on the other hand, the IHC test is quite expensive. Patients often do an H&E first, and then they might perform an IHC to confirm it. As there are multiple stages of cancer, they have to do this quite regularly to get to know the current stage of cancer. Patients will be able to determine the cancer's current stage using this research work and whatever test findings (IHC or H&E) they are comfortable with.

### B. Data Pre-Processing

The pre-processing step is essential for removing various types of noise from tissue images. Similarly, we preprocess our dataset for better accuracy. The dataset contains images of HE and IHC staining. We began by labeling our data set. Our dataset contains four categories of images that correspond to the four stages of HER2 breast cancer. So, we label the dataset with the HER2 breast cancer stages. Moreover, we resized our dataset. The data set contains high-resolution images. Therefore, resizing is required for both the training and compatibility of our model. Our images were resized to 256 × 256 pixels. The next thing we do is divide our dataset into a train and test set. For HE, there are 3896 training images and 977 test images. For IHC, there are 3896 training images and 977 test images. Both HE and IHC data samples were obtained from the same patients. The HER2+ status of both HE and IHC at the same level remains the same. So, from the dataset, we know what it looks like for the same patient's medical test images for both IHC and HE.

### C. Model Implementation

The performance of the neural network will improve as the depth of the network is increased, but this progress will come at the cost of time and processing resources. Deep learning that employs transfer learning (TL) therefore seems as a strategy to reduce training expenses. In a nutshell, initiating a newly developed model's weights according to an existing framework's learnt weights to aid training is called transfer learning. Because certain information is important, the new model's training performance may be sped and enhanced by sharing the parameters of the trained model with the new model via TL, rather than beginning from blank, as we did in the earlier phase of our work (CNN). The need for data has been increased as par the popularity of deep learning in recent years, therefore TL has progressed.

The fundamental architecture [29] of Inception-v3, an expanded version of the well-known GoogLeNet that has efficiently classified images via transfer learning, is shown in Figure 7. Inception-v3 introduced an inception model that concatenates many different-sized convolutional filters into a new filter, following GoogLeNet. Such a design minimizes the complexity of the calculation by lowering the number of parameters that must be trained. The Inception V3 model has 42 layers, which is slightly more than the V1 and V2 models. However, the efficiency of this model is truly remarkable. Because of this, this research uses the Inception V3 model. Usually, the Inception module includes three distinct convolution sizes and one maximum pooling. After the convolution operation, the channel is gathered for the previous layer's network output, and then nonlinear fusion is performed. Within this interpretation, the network's representation and learnability for multiple aspects can be enhanced, and overfitting can be avoided. Inception-V3 is primarily a network structure created by Keras and pre-trained with Image Net. The default image size for input is 256 × 256 pixels with three channels. This research trained the dataset with the original Inception V3 model, but the validation accuracy and loss are not mentionable compared to other state-of-the-art works. One of our research objectives is to tune the Inception V3 model so that we can get better accuracy for present work. To achieve that, we have to add an additional layer and different parameter values. The first parameter is the addition of regularization after every hidden layer.

For regularization, a batch normalization (BN) layer is introduced in between the auxiliary classifier and the fully connected (FC) layer in the Inception-v3. The batch gradient descent approach can be used to accelerate the training speed and model convergence of a deep neural network in the BN model. The BN formulae are specified as follows:

$$B = \{X_{1...m}\}, \gamma, \beta$$
$$\{y_i = BN_{\gamma,\beta}(X_i)\}$$

$$\mu_B \leftarrow \frac{1}{m}\sum_{i=1}^{m} X_i \quad \text{......(i)}$$

$$\sigma_B^2 \leftarrow \frac{1}{m}\sum_{i=1}^{m}((X_i - \mu_B))^2$$

$$\hat{X}_i \leftarrow \frac{X_i - \mu_B}{\sqrt{\sigma_B^2 + \varepsilon}}$$

$$y_i \leftarrow \gamma \hat{X}_i + \beta = BN_{\gamma,\beta}(X_i) \ldots (ii)$$

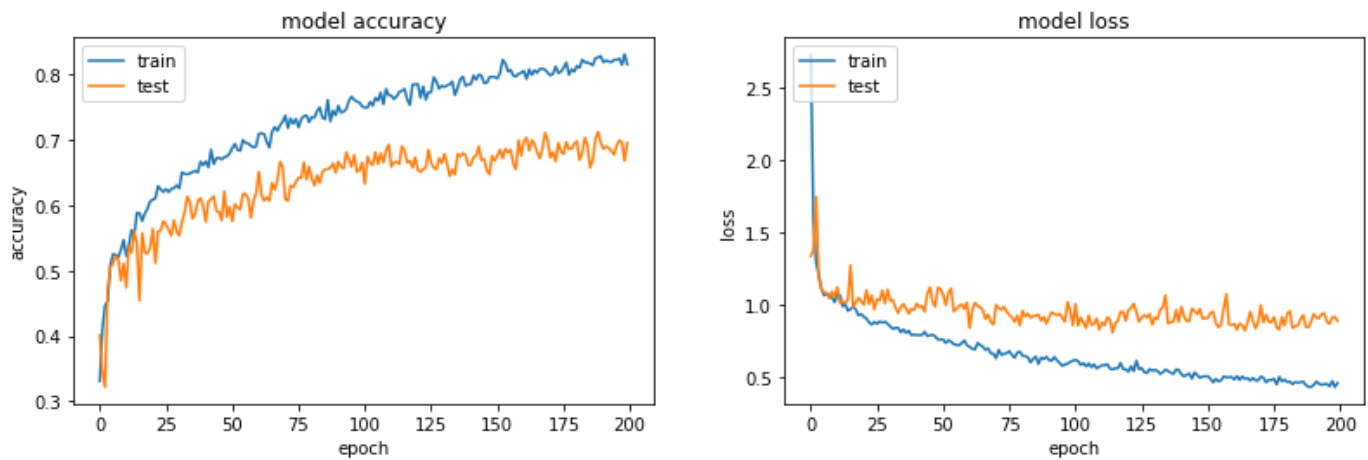

Fig. 5. Left side graph for the performance evaluation Accuracy and Right-side graph for the performance evaluation Loss on H&E Dataset.

The average value in one dimension is denoted by µB, the standard deviation across all feature map dimensions is denoted by σB2, and the constant is learned. Batch B's minimum activation value is denoted by x, and the number of activation values is denoted by m. Parameters can be learned to adjust the variance and average value distributions, respectively, γ and β.

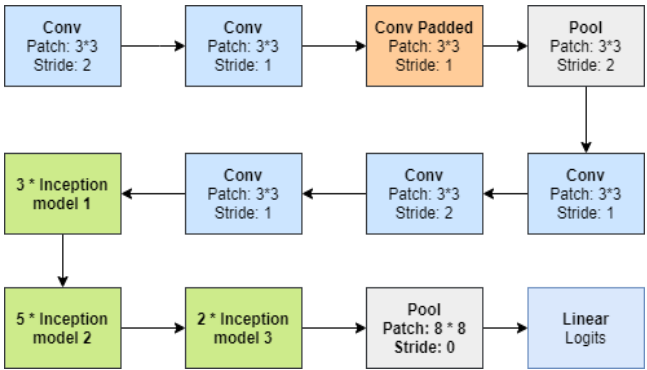

Figure 7: Original Inception V3 Architecture.

The research utilizes the "ConvoHER2" model. This CNN model is comprised of a pre-trained Inception-V3 model and adds some additional layers to it to achieve the highest level of accuracy. Features extraction was done by the Inception-V3 module. It's pre-trained model, which has been trained on over 1.4 million photos and over 1,000 classes. As we have already discussed that the Inception-V3 architecture utilizes CNN to extract features, which acts as an image classifier. In our research, 80% of the data was kept as training data and 20% was kept as testing data. In our transfer learning method, we keep the parameters of the previous layer and remove the last layer of the Inception-V3 model, then retrain the last layer. The final layer has the same number of output nodes as the number of categories in the dataset. We utilized a weight that had already been learned using ImageNet. The customized, fully linked layer was trained. In all, there are 9,724,932 trainable parameters. The Adam optimizer is used in the experiment, with categorical cross-entropy as the loss function, 256 batches, and a learning rate of 1e-4, while the input tensor of InceptionV3 is set to (256, 256, 3). Rectified linear unit (ReLu) for the deep layer and Softmax for the output layer was chosen as activation functions. For all positive values, ReLu remains linear in terms of positive data, and zero for any negative data. This simple calculation resulted in reduced training time. To mitigate vanishing gradient problem ReLU has been utilized over activation functions. It can be stated mathematically as indicated in eq (iii):

$$X(x) = max(0, x) \ldots (iii)$$

X () is the function, 0 is the initial value, x is the input, and a maximum of 0 is considered. Due to the fact that the ReLU function returns 0 for all negative integers, the starting value is set to 0.

We are predicting multiple classes in our work. That's why we have used the Softmax activation function in the output layer. Mathematically, it can be represented as shown in eq (iv), where z is the value of neurons from the output layer.

$$\text{softmax}(z_i) = \frac{exp(z_j)}{\sum_j exp(z_j)} \quad \ldots\ldots(iv)$$

IV. EXPERIMENTAL RESULTS

We have got the accuracy for the original inception V3, which is 76% on average. During the training period, we tweaked a variety of variables. That was covered in the methods section, previously. We first added two more layers and one output layer to the model. To be on the safe side, we have set a learning rate of 1e-4, which will optimize the model with a local minimum while learning slowly. This study has also included a few additional standard parameters, such as batch size of 256 and an 80-20% split of the image dataset. However, after adding the batch normalization after each layer and enhancing the dataset with various angles, sizes, etc., the outcome considerably improved. We have used the keras library known as "Keras-Tunner" to discover the quantity of layers and nodes that may result in the greatest accuracy. Although it didn't provide the precise layers or nodes we used in our work, it did provide a foundation for the model that allowed us to start getting decent accuracy and gradually enhance it over time with a few node and parameter adjustments which is provided on Table I.

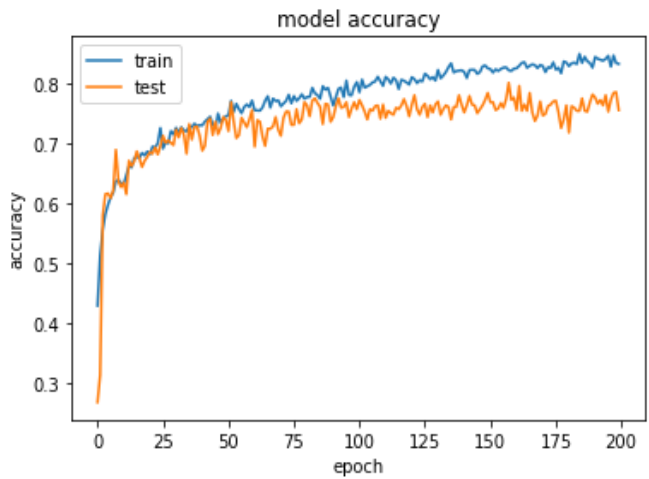 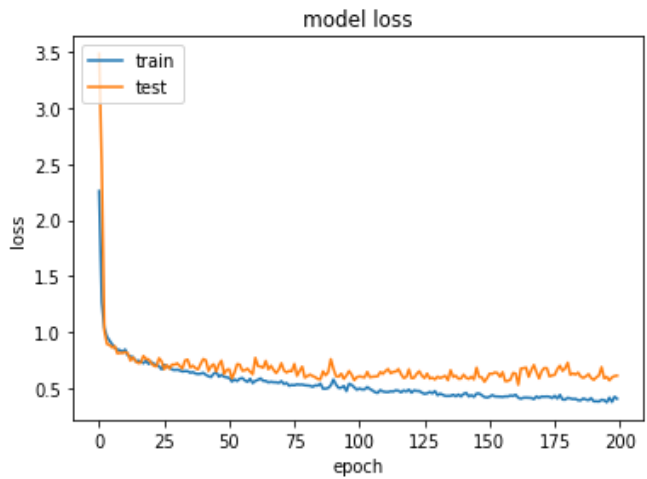

Fig. 6. Left side graph for the performance evaluation Accuracy and Right-side graph for the performance evaluation Loss on IHC Dataset.

With an input shape of, the basic model Inception-V3 for transfer learning with pre-trained weights of image-net is used (256,256,3). Three additional dense layers are created during the classification step, with the first layer having 2048 units. ReLu along with BN acts as an activation function. The activation function of ReLu with BN is examined for both layers in the second layer and third layer 1536 units. The optimizer under consideration is ADAM, which converges quicker. Categorical cross-entropy is used for the accuracy metrics and loss function. The batch size for the training procedure is 256, and the total number of images trained is 3896. For each iteration/epoch, training logs are created, and loss is tracked. The current loss is compared to the prior loss, and then the model is stored. We've trained the model for 200 epochs in our work.

Fig. 5 and 6 illustrates the performance evaluation of the Inception-V3 architecture on the HE & IHC dataset respectively with respect to accuracy and loss while training and validation.

Our research uses 80 percent of the images from the dataset for the training phase, resulting in training accuracy of 85.10 percent for H&E and 87.79 percent for IHC, respectively. This implies that out of 3896 images, the model has fitted successfully at 3233 images and 3350 images, respectively. It's clear that several photographs with equivalent texture, shape, and color were misclassified; otherwise, it delivered effective result with state-of-the-art accuracy. Table II displays how much better accuracy this research paper's model achieves compared to other state-of-the-art deep learning models such as DenseNet, HASHI algorithm, etc.

## V. CONCLUSION

The research begins with a discussion of the significance of H&E and IHC imaging investigations in determining the present stage of HER2. The data for this study is separated into four groups, ranging from zero to three. The customized convolution neural network-based model has been trained on almost 4,000 H&E and IHC images, which are individually at the top of the imageNet weights, enhancing our model's capacity to swiftly detect essential characteristics. Furthermore, in our study, we suggested a unique CNN model that was more accurate. Model accuracy during the training phase is 88 percent for IHC and 85 percent for HE images at 200 epochs. The future scope of our study activity is limitless as medical experts uncover new breakthroughs to identify and treat breast cancer. The next step in this research will be to segment the pictures to see which areas of the HE and IHC test samples are the most damaged by HER2. Furthermore, by incorporating the XAI, anybody can use a deep learning model to identify which area or region of the body is most afflicted by breast cancer. Additionally, we are unable to access the high pixel dataset on our system for this research project. The datasets used in cancer imaging pictures is sometimes difficult to acquire and utilize with our research tools. In order to determine the correctness of the ConvoHER2 model's results, additional high-quality datasets will be used in this paper's future work.

TABLE II
PERFORMANCE COMPARISON OF THE PROPOSED MODEL WITH RECENT EXISTING WORKS

| Authors | Dataset | Model | Accuracy |
| --- | --- | --- | --- |
| J. Zhou et al [19] | MRI images | SVM | 79.5% |
| Z. Xu et al.[20] | Ultrasound images | DenseNet | 80.56% |
| Oliveira et al. [30] | HER2SC | HASHI algorithm | 83.3% |
| In this Paper | BCI Dataset (H&E) | convoHER2 | **85.1%** |
| | BCI Dataset (IHC) | convoHER2 | **87.79%** |

## REFERENCES

[1] Sharma, Rajesh. "Breast cancer incidence, mortality and mortality-to-incidence ratio (MIR) are associated with human development, 1990–2016: evidence from Global Burden of Disease Study 2016." *Breast Cancer* 26, no. 4 (2019): 428-445.
[2] Siegel, Rebecca L., Kimberly D. Miller, and Ahmedin Jemal. "Cancer statistics, 2016." *CA: a cancer journal for clinicians* 66, no. 1 (2016): 7-30.

[3] Dai, Xiaofeng, Ting Li, Zhonghu Bai, Yankun Yang, Xiuxia Liu, Jinling Zhan, and Bozhi Shi. "Breast cancer intrinsic subtype classification, clinical use and future trends." *American journal of cancer research* 5, no. 10 (2015): 2929.

[4] Zhu, Chuang, Fangzhou Song, Ying Wang, Huihui Dong, Yao Guo, and Jun Liu. "Breast cancer histopathology image classification through assembling multiple compact CNNs." *BMC medical informatics and decision making* 19, no. 1 (2019): 1-17.

[5] Wolff, Antonio C., M. Elizabeth Hale Hammond, Kimberly H. Allison, Brittany E. Harvey, Pamela B. Mangu, John MS Bartlett, Michael Bilous et al. "Human epidermal growth factor receptor 2 testing in breast cancer: American Society of Clinical Oncology/College of American Pathologists clinical practice guideline focused update." *Archives of pathology & laboratory medicine* 142, no. 11 (2018): 1364-1382.

[6] La Barbera, David, António Polónia, Kevin Roitero, Eduardo Conde-Sousa, and Vincenzo Della Mea. "Detection of her2 from haematoxylin-eosin slides through a cascade of deep learning classifiers via multi-instance learning." *Journal of Imaging* 6, no. 9 (2020): 82.

[7] Zaha, Dana Carmen. "Significance of immunohistochemistry in breast cancer." *World journal of clinical oncology* 5, no. 3 (2014): 382.

[8] Khameneh, Fariba Damband, Salar Razavi, and Mustafa Kamasak. "Automated segmentation of cell membranes to evaluate HER2 status in whole slide images using a modified deep learning network." *Computers in biology and medicine* 110 (2019): 164-174.

[9] Nahta, Rita, Dihua Yu, Mien-Chie Hung, Gabriel N. Hortobagyi, and Francisco J. Esteva. "Mechanisms of disease: understanding resistance to HER2-targeted therapy in human breast cancer." *Nature clinical practice Oncology* 3, no. 5 (2006): 269-280.

[10] Yarden, Yosef. "Biology of HER2 and its importance in breast cancer." *Oncology* 61, no. Suppl. 2 (2001): 1-13.

[11] Romond, Edward H., Edith A. Perez, John Bryant, Vera J. Suman, Charles E. Geyer Jr, Nancy E. Davidson, Elizabeth Tan-Chiu et al. "Trastuzumab plus adjuvant chemotherapy for operable HER2-positive breast cancer." *New England journal of medicine* 353, no. 16 (2005): 1673-1684.

[12] Wang, Dayong, Aditya Khosla, Rishab Gargeya, Humayun Irshad, and Andrew H. Beck. "Deep learning for identifying metastatic breast cancer." *arXiv preprint arXiv: 1606.05718* (2016).

[13] Ravì, Daniele, Charence Wong, Fani Deligianni, Melissa Berthelot, Javier Andreu-Perez, Benny Lo, and Guang-Zhong Yang. "Deep learning for health informatics." *IEEE journal of biomedical and health informatics* 21, no. 1 (2016): 4-21.

[14] Litjens, Geert, Thijs Kooi, Babak Ehteshami Bejnordi, Arnaud Arindra Adiyoso Setio, Francesco Ciompi, Mohsen Ghafoorian, Jeroen Awm Van Der Laak, Bram Van Ginneken, and Clara I. Sánchez. "A survey on deep learning in medical image analysis." *Medical image analysis* 42 (2017): 60-88.

[15] Shen, Dinggang, Guorong Wu, and Heung-Il Suk. "Deep learning in medical image analysis." *Annual review of biomedical engineering* 19 (2017): 221.

[16] Liu, Shengjie, Chuang Zhu, Feng Xu, Xinyu Jia, Zhongyue Shi, and Mulan Jin. "BCI: Breast Cancer Immunohistochemical Image Generation through Pyramid Pix2pix." In *Proceedings of the IEEE/CVF Conference on Computer Vision and Pattern Recognition*, pp. 1815-1824. 2022.

[17] G. Modak, S. S. Das, M. A. Islam Miraj and M. K. Morol, "A Deep Learning Framework to Reconstruct Face under Mask," 2022 7th International Conference on Data Science and Machine Learning Applications (CDMA), 2022, pp. 200-205, doi: 10.1109/CDMA54072.2022.00038.

[18] S. Paik et al., "Real-World Performance of HER2 Testing--National Surgical Adjuvant Breast and Bowel Project Experience", *JNCI Journal of the National Cancer Institute*, vol. 94, no. 11, pp. 852-854, 2002. Available: 10.1093/jnci/94.11.852.

[19] J. Zhou et al., "Radiomics Signatures Based on Multiparametric MRI for the Preoperative Prediction of the HER2 Status of Patients with Breast Cancer", *Academic Radiology*, vol. 28, no. 10, pp. 1352-1360, 2021. Available: 10.1016/j.acra.2020.05.040.

[20] Z. Xu et al., "Predicting HER2 Status in Breast Cancer on Ultrasound Images Using Deep Learning Method", *Frontiers in Oncology*, vol. 12, 2022. Available: 10.3389/fonc.2022.829041.

[21] D. Anand et al., "Deep Learning to Estimate Human Epidermal Growth Factor Receptor 2 Status from Hematoxylin and Eosin-Stained Breast Tissue Images", *Journal of Pathology Informatics*, vol. 11, no. 1, p. 19, 2020. Available: 10.4103/jpi.jpi_10_20 [Accessed 8 June 2022].

[22] S. Farahmand et al., "Deep learning trained on H&E tumor ROIs predicts HER2 status and Trastuzumab treatment response in HER2+ breast cancer", 2021. Available: 10.1101/2021.06.14.448356 [Accessed 9 June 2022].

[23] W. Lu, M. Toss, M. Dawood, E. Rakha, N. Rajpoot, and F. Minhas, "SlideGraph+: Whole slide image level graphs to predict HER2 status in breast cancer," *Medical Image Analysis*, vol. 80, p. 102486, Aug. 2022, doi: 10.1016/j.media.2022.102486.

[24] C. Cordeiro, S. Ioshii, J. Alves and L. Oliveira, "An Automatic Patch-based Approach for HER-2 Scoring in Immunohistochemical Breast Cancer Images Using Color Features", *arXiv.org*, 2018. [Online]. Available: https://doi.org/10.48550/arXiv.1805.05392. [Accessed: 08- Jun-2022].

[25] R. Mukundan, "Analysis of Image Feature Characteristics for Automated Scoring of HER2 in Histology Slides," *Journal of Imaging*, vol. 5, no. 3, p. 35, Mar. 2019, doi: 10.3390/jimaging5030035.

[26] M. Saha and C. Chakraborty, "Her2Net: A Deep Framework for Semantic Segmentation and Classification of Cell Membranes and Nuclei in Breast Cancer Evaluation," *IEEE Transactions on Image Processing*, vol. 27, no. 5, pp. 2189–2200, 2018, doi: 10.1109/TIP.2018.2795742.

[27] S. Tewary and S. Mukhopadhyay, "HER2 Molecular Marker Scoring Using Transfer Learning and Decision Level Fusion", *Journal of Digital Imaging*, 2021. Available: 10.1007/s10278-021-00442-5 [Accessed 10 June 2022].

[28] T. Qaiser et al., "HER2 challenge contest: a detailed assessment of automated HER2 scoring algorithms in whole slide images of breast cancer tissues", *Histopathology*, vol. 72, no. 2, pp. 227-238, 2017. Available: 10.1111/his.13333.

[29] Nguyen, Long & Lin, Dongyun & Lin, Zhiping & Cao, Jiuwen. (2018). Deep CNNs for microscopic image classification by exploiting transfer learning and feature concatenation. 1-5. 10.1109/ISCAS.2018.8351550.

[30] L. Alzubaidi *et al.*, "Review of deep learning: concepts, CNN architectures, challenges, applications, future directions," *Journal of Big Data*, vol. 8, no. 1, Mar. 2021, doi: 10.1186/s40537-021-00444-8.

[31] Oliveira, Sara P., João Ribeiro Pinto, Tiago Gonçalves, Rita Canas-Marques, Maria-João Cardoso, Hélder P. Oliveira, and Jaime S. Cardoso. "Weakly-supervised classification of HER2 expression in breast cancer hematoxylin and eosin-stained slides." *Applied Sciences* 10, no. 14 (2020): 4728.

[32] A. M. Faruk, H. A. Faraby, M. Azad, M. Fedous, and M. Morol, "Image to Bengali caption generation using deep CNN and bidirectional gated recurrent unit," 2020, https://arxiv.org/abs/2012.12139.

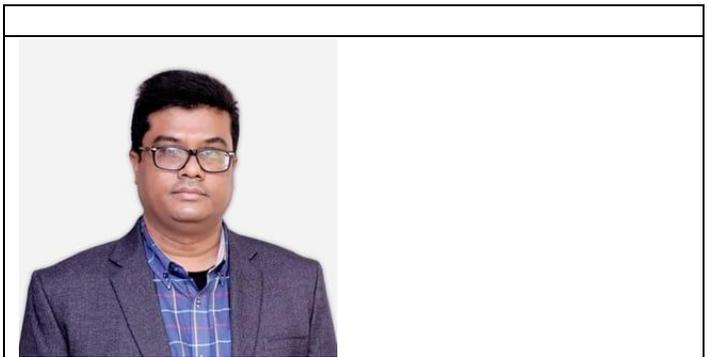

**Dr. M. F. MRIDHA** (Senior Member, IEEE) received the Ph.D. degree in AI/ML from Jahangirnagar University, in

2017.He is currently working as an Associate Professor in the Department of Computer Science, American International University-Bangladesh.His research experience, within both academia and industry, results in over 120 journal and conference publications. His research interests include artificial intelligence (AI), machine learning, deep learning, natural language processing (NLP), and big data analysis. He has served as a program committee member for several international conferences/workshops. He has also served as an associate editor for several journals.

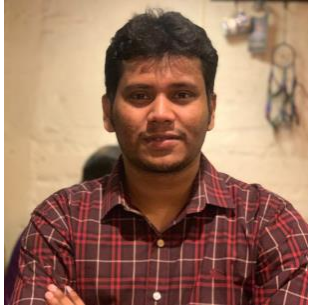

**Md. Kishor Morol** received his B.Sc. and M.Sc. Degrees in Computer Science from American International University-Bangladesh (AIUB) in December 2016 and December 2018, respectively. Currently He is working as an Assistant Professor in Computer Science department at AIUB. His research interests are AppliedMachine Learning, Natural Language Processing, Deep learning and Data and Medical image processing

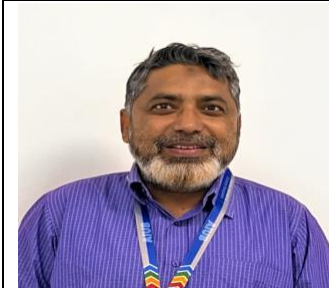

**Prof. Dr. Md. Asraf Ali** is currently working in the Department of Computer Science at American International University-Bangladesh (AIUB). He is actively involved in research in the field of Artificial Intelligence of Biological Signal Processing, Bioinformatics, and Machine Learning. He also provides several voluntary tasks to the research community as reviewer and evaluating PhD thesis as foreign external examiner in different countries. Prior to start his professional work, he achieved his academic degrees BSc & MSc in Computer Science from University of Madras, India and PhD in Biomedical Electronic Engineering form Universiti Malaysia Perlis (UniMAP), Malaysia.

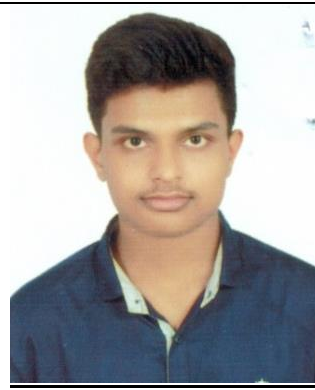

**Md Sakib Hossain Shovon** has recently finished his BSc In Computer Science & Engineering-Major in Information Systems at American International University-Bangladesh AIUB in Computer Science and Engineering. His research interest is in artificial intelligence (AI), machine learning (NL), deep learning(DL), and Computer Vision, especially in Medical Imaging.